\documentclass{elsart}

\usepackage{natbib}

\usepackage{amssymb}

\usepackage{amsfonts}
\usepackage{graphicx}% Include figure files
\usepackage{bm}% bold math

\def\const{\mathop {\rm const}\nolimits } % const
\def\sn{\mathop {\rm sn}\nolimits } % Jacobi's sine
\def\dn{\mathop {\rm dn}\nolimits } % Jacobi's delta
\def\nd{\mathop {\rm nd}\nolimits } % Jacobi's delta
\def\cd{\mathop {\rm cd}\nolimits } % Jacobi's delta
\def\K{\mbox{\rm K}} % Elliptic integral of 1-st kind
\def\E{\mbox{\rm E}} % Elliptic integral of 2-nd kind

\bibpunct{(}{)}{;}{a}{,}{,}

\begin{document}

\begin{frontmatter}

\title{Cascade unlooping of a low-pitch helical spring under tension}

\author{E.~L.~Starostin, G.~H.~M.~van~der~Heijden}
\ead{e.starostin@ucl.ac.uk, g.heijden@ucl.ac.uk}

\address{Centre for Nonlinear Dynamics, University College London,\\
 Gower Street, London WC1E 6BT, UK}

\date{\today}

\begin{abstract}
We study the force vs extension behaviour of a helical spring made of a thin torsionally-stiff anisotropic elastic rod.
Our focus is on springs of very low helical pitch. For certain parameters of the problem such a spring is found not to unwind when pulled but rather to form hockles that pop-out one by one and lead to a highly non-monotonic force-extension curve. Between abrupt loop pop-outs this curve is well described by the planar elastica whose relevant solutions are classified. Our results may be relevant for tightly coiled nanosprings in future micro- and nano(electro)mechanical devices.

\end{abstract}

\begin{keyword}
thin anisotropic elastic rod \sep helical spring \sep cascade unlooping \sep planar elastica \sep force-extension response \sep loop pop-out \sep hysteresis

\end{keyword}

\end{frontmatter}

% main text

\section{Introduction}
\begin{figure}% FIGURE ================================================
\vspace{0.5cm}
\centering
\includegraphics[height=0.9in]{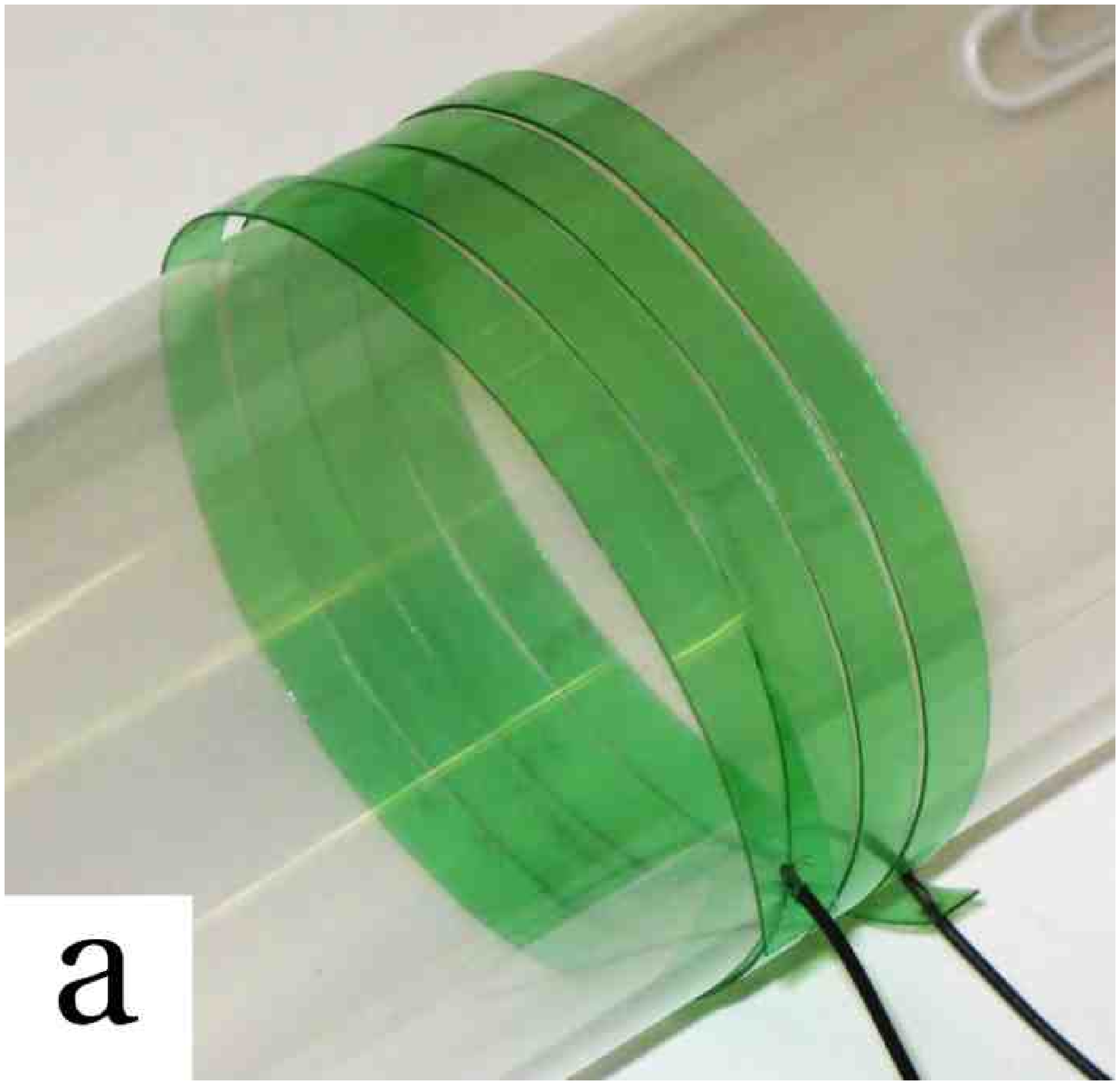}
\includegraphics[height=0.9in]{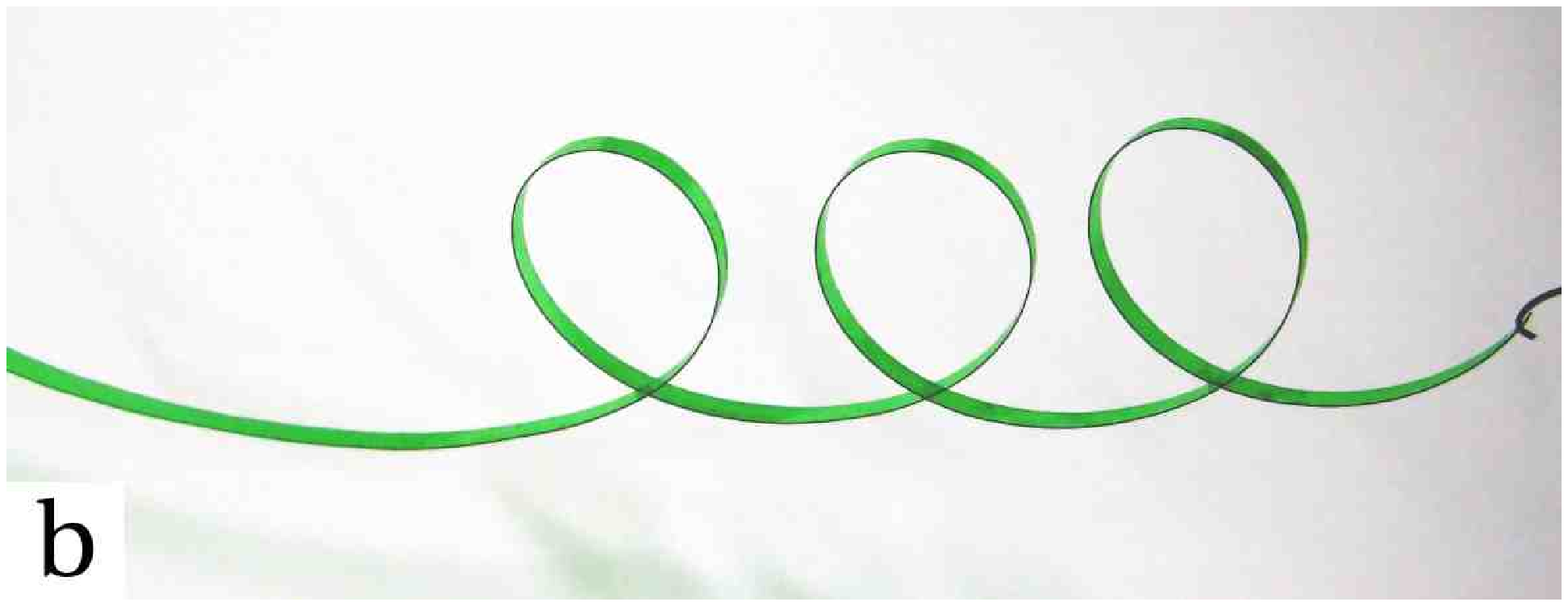}\\
\includegraphics[width=2.6in]{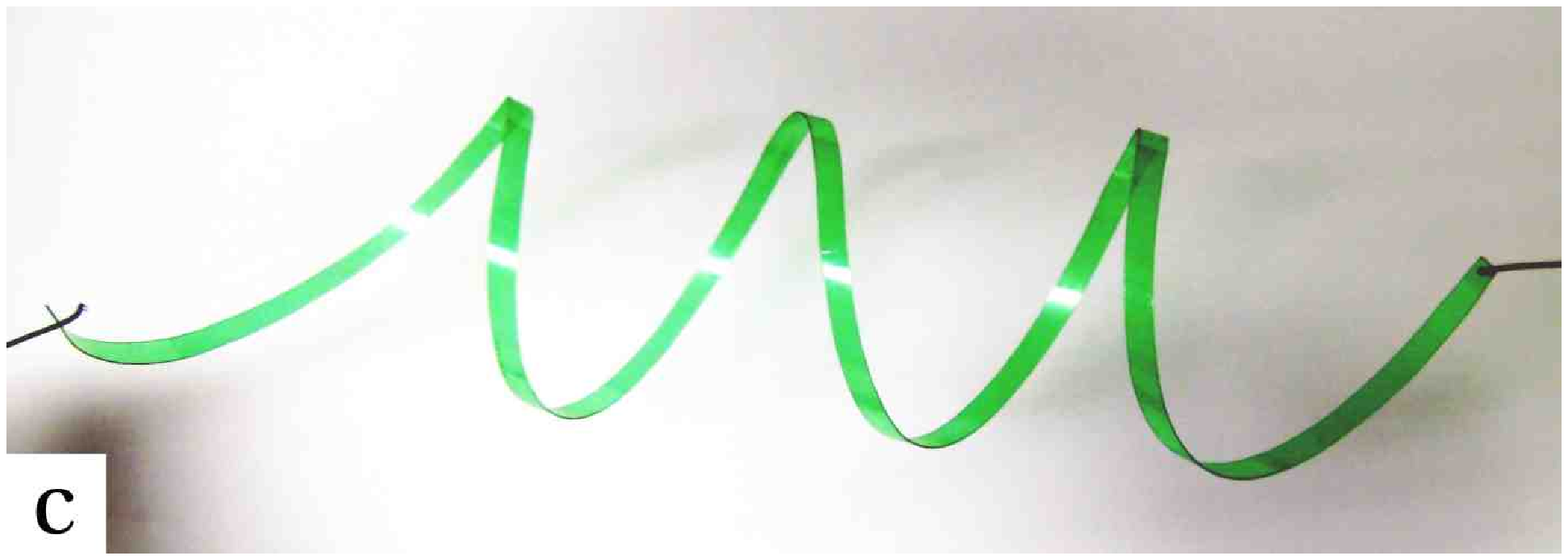}
\includegraphics[width=2.6in]{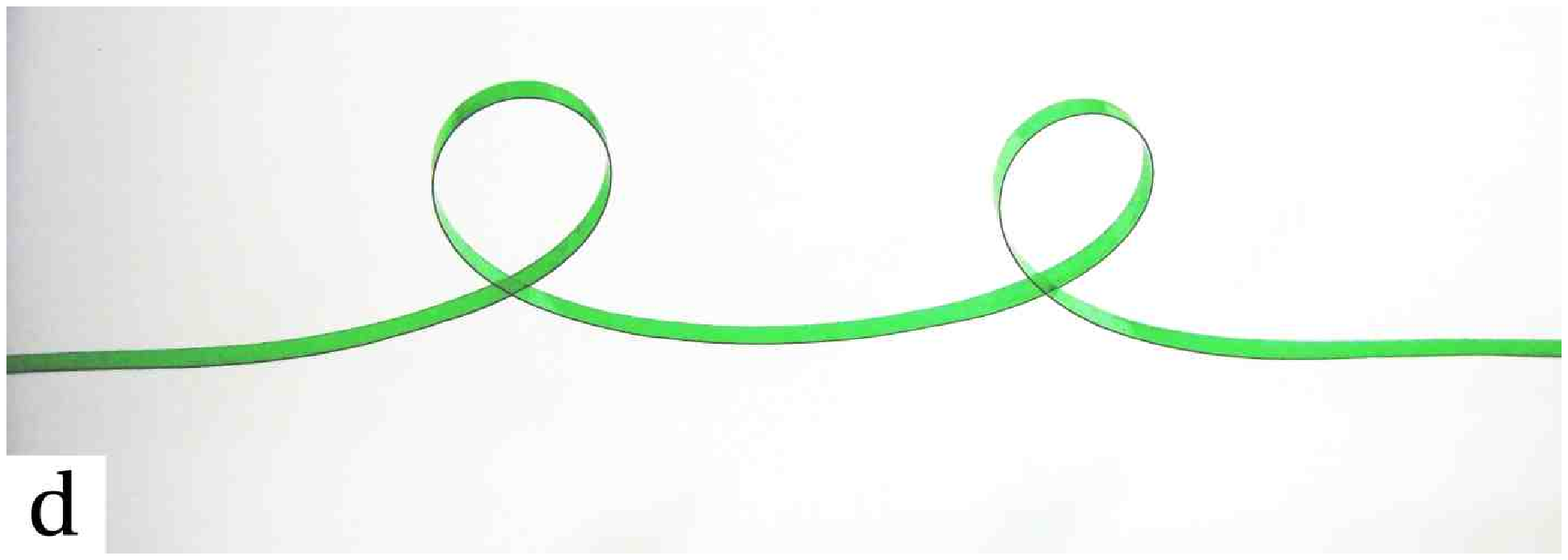}
\includegraphics[width=2.6in]{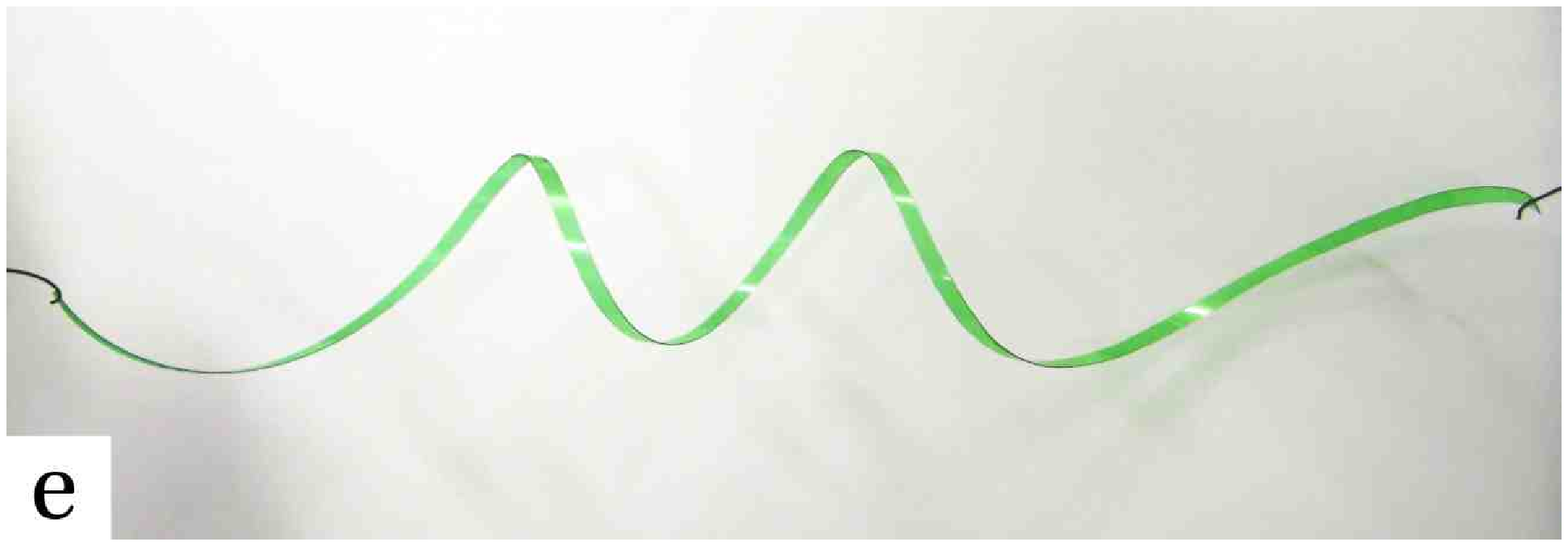}
\includegraphics[width=2.6in]{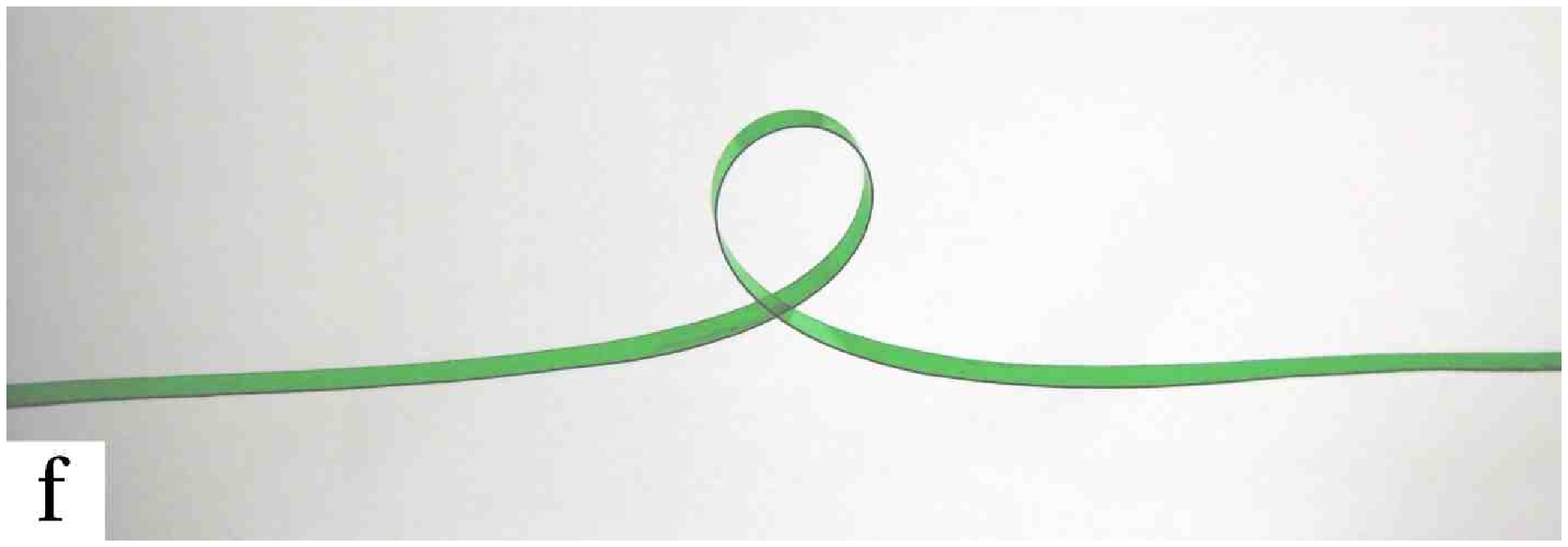}
\includegraphics[width=2.6in]{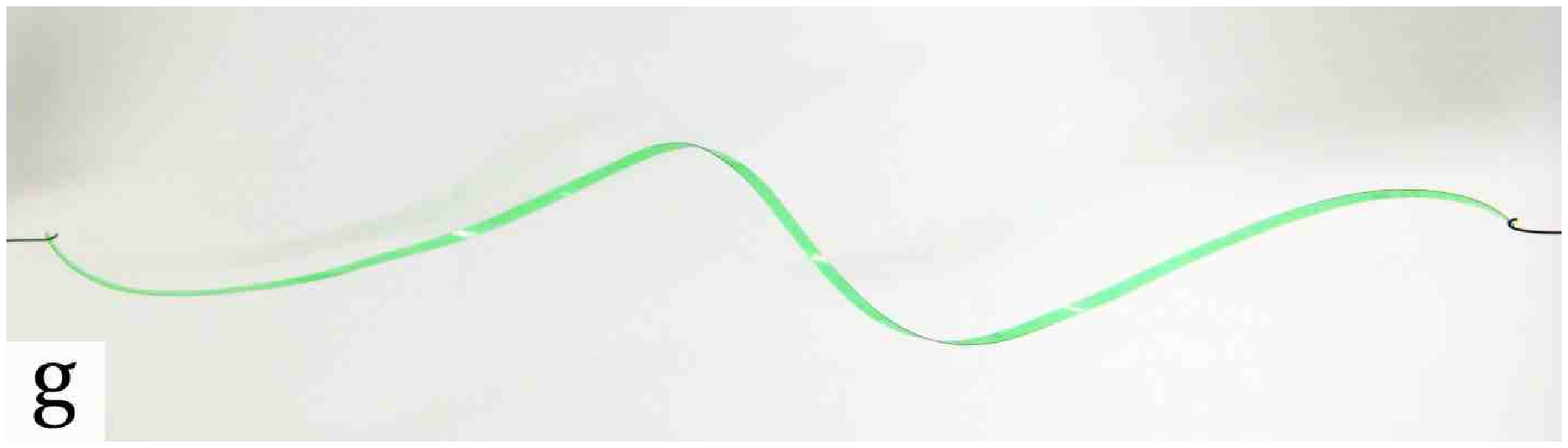}
\includegraphics[width=2.6in]{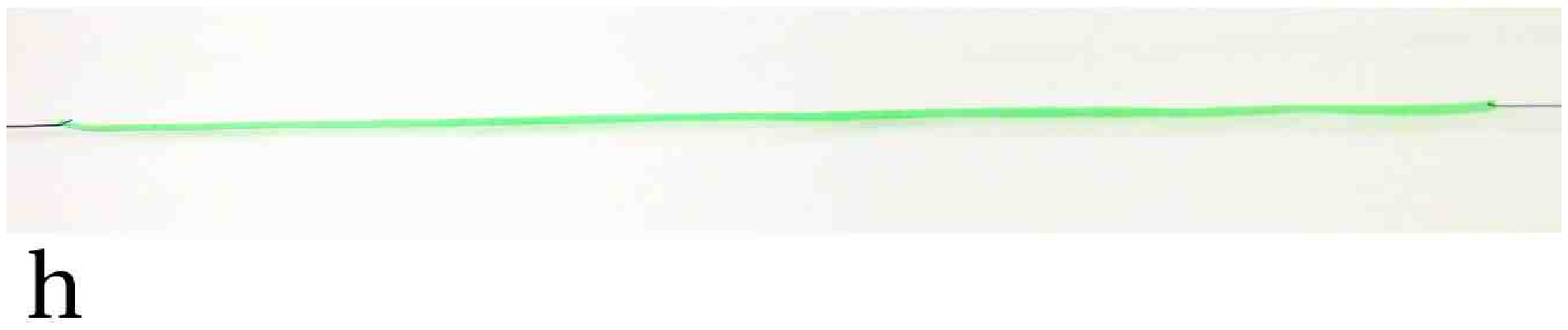}
\caption{Photographs of tensile tests on a plastic helical strip of small pitch. The unstressed strip has four loops, which are successively lost as the force is increased.}
\label{photos}
\end{figure}

%\label{}
Take an elastic rod or strip that is coiled when left free and pull the ends apart without fixing their orientation. We can do this by attaching strings to the ends and applying a tensile force to the strings (see Fig.~\ref{photos}).
If we imagine the rod to be held between two parallel glass plates, forcing it to adopt an essentially planar configuration, then we would observe the formation of a number of loops given by the number of initial coils. The loops have nowhere to go under increased stretching of the rod and grow into tight hockles with a highly localised curvature. Without the constraining walls the planar configuration becomes unstable at a certain force and unlooping occurs. However, this unlooping force can be very high, especially if the rod has a flat cross-section, and the structure may get damaged (e.g., through plastic deformation) before loop pop-out occurs. Indeed, hockles (or kinks or snarls) are a serious concern not only in the use of garden hoses but also in such slender industrial structures as marine pipelines, mooring ropes, textile yarns and communication cables \citep{Coyne90}. Pipes, ropes, yarns and cables often acquire intrinsic curvature due to creep from storage on a drum, putting them at risk for the above hockling scenario.

Helical springs have also recently been fabricated at the micro and nano scale
\citep{Gao05,Cho06,Prinz06}, where they hold great promise as essential parts of future micro- or nano(electro)mechanical devices such as chemical or biological sensors, cantilevers, resonators, inductors, actuators, etc.~\citep{Bell06a}.
Nanohelices have been synthesised from different materials leading to structures with unusual mechanical properties, and force-extension experiments on nanosprings, nanocoils and nanobelts are actively being performed \citep{Chen03,Kratochvil07}. For instance, ZnO and Si$_3$N$_4$ springs have been shown to be superelastic, i.e., they can be stretched to an almost straight line and when released they restore their shape without damage \citep{Gao06,Cao08}. An interesting development is the fabrication of nanosprings with very small pitch \citep{Zhang06}. Such springs are particularly desirable because they allow for a large magnetic flux density.

\begin{figure}% FIGURE ================================================
\begin{center}
\includegraphics[width=3in]{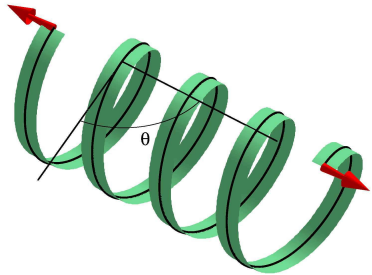}
\caption{\label{fig:spring} Schematic diagram of a pulled spring of helical angle $\theta$. A flat strip is drawn to emphasise the anisotropic bending stiffness of the rod.}
\end{center}
\end{figure}

In this paper we consider the problem of successive hockling and unlooping of a helically coiled strip of low pitch under tensile load. This problem was briefly touched upon in \citep{Starostin08a}, where we modelled the strip as an inextensible helical shell motivated by multistability observed in cholesterol ribbons \citep{Smith01}. Here we do not make the inextensibility assumption and model the strip in the more common way as a transversely anisotropic rod \citep{Zhou05}.
The undeformed rod is curved about the axis of least bending stiffness (so that our spring is close to a clock spring, rather than, say, a slinky toy; see Fig.~\ref{fig:spring}).
It is worth pointing out here that although the rod will be assumed to have an inextensbile centreline, this rod model is not equivalent to the inextensible strip model in the limit of infinitesimal width, as shown by \citet{Starostin07}. The two models apply to different circumstances.

We choose zero-moment boundary conditions because these seem to be a reasonable choice to avoid kinking as the rod is free to unwind (the loops are not `locked in' as in the case of clamped boundary conditions). Nevertheless, we show that under certain conditions on the elastic and geometrical parameters a tightly coiled strip when pulled behaves in a highly nonlinear way. Under small forces the spring's force-extension behaviour is Hookean, but at larger extensions the loading curve develops hysteresis cycles as more and more loops pop out. These pop-outs form brief and abrupt three-dimensional excursions in a process that is otherwise remarkably well approximated by planar elastic behaviour.

This planar behaviour is described by the Euler elastica subject to a constant intrinsic curvature. The fact that a low-pitch helical spring is found to jump between planar states justifies a revisit of the Euler elastica and we give what appears to be a new classification of its solutions subject to zero-moment end conditions. The corresponding force-extension curves provide the scaffold of the tensile response of the helical spring. At larger pitch no tight hockles occur,
although multistability may be observed \citep{Kessler03}.

Multi-loop hockling and unlooping does not seem to have been studied before.
\cite{Coyne90} only considered a single loop. \cite{Yabuta82} discussed cable loop stability assuming a single tightly coiled helical turn. Both works assume the unstressed rod to be straight and avoid boundary conditions by assuming the cable to be infinitely long. We model the strip as a finite-length intrinsically curved elastic rod of anisotropic (flat) cross-section.

The organisation of the paper is as follows. In Section 2 we set up the equilibrium equations for elastic rod statics. In Section 3 we classify the planar solutions of this strip and find the force-extension curves which the curves for a coiled strip will be compared against (in Section 4). The main result of the paper is Fig.~\ref{spring}, which shows the hysteretic force-extension behaviour characterising multi-loop pop-out. We briefly contrast this behaviour with that of springs of different geometric and elastic parameters. Section 5 concludes the paper.

%%%%%%%%%%%%%%%%%%%%%%%%%%%%%%%%%%%%%%%%%%%%%%%%%%%%%%%%%%%%%%%%%%%%%%%%%%%%%%%%%%%%%%%%%%%%%%%%%%%%%%%%%%%%%%%%%%%%%%%%%%%%
\section{Anisotropic rod model}

We consider a thin inextensible elastic rod with non-circular cross-section and free from distributed external loads. Let the centreline of the rod be $\bm{r}(s)$, where $s \in [0,1]$ is arclength and the length of the rod has been scaled to 1. 

The two principal axes of the cross-section and the tangent to the centreline, $\bm{t}=\bm{r}'$, constitute an orthonormal material frame $\{\bm{d}_1,\bm{d}_2,\bm{d}_3\}$, where $\bm{d}_3=\bm{t}$ and a prime denotes differentiation with respect to $s$. By orthonormality we have the kinematic equations
\begin{equation}
\bm{d}_i'=\bm{u}\times\bm{d}_i, \quad\quad i=1,2,3,
\end{equation}
where $\bm{u}=(u_1, u_2, u_3)^T$ is the strain vector written in the material frame.

Let $\bm{F}(s)$ and $\bm{M}(s)$ be the resultant internal force and moment that the material of $[s,s_1)$ in the rod exerts on the material of $(s_0,s]$, where $s_0<s<s_1$. In case there are no distributed loads the force and moment balance equations take the form \citep{Love27}
\begin{equation}
\bm{F}'=\bm{0}, \quad\quad \bm{M}'+\bm{t}\times\bm{F}=\bm{0}.
\label{eq:bal}
\end{equation}
We close the system of equations by specifying (linear) constitutive relations
\begin{eqnarray}
&& \bm{M}\cdot\bm{d}_1=A(u_1-u_{10}), \nonumber \\
&& \bm{M}\cdot\bm{d}_2=B(u_2-u_{20}), \label{eq:constit} \\
&& \bm{M}\cdot\bm{d}_3=C(u_3-u_{30}), \nonumber
\end{eqnarray}
where we have allowed for non-zero curvature and torsion of the rod in its unstressed state. Here $A$ and $B$ are the principal bending stiffnesses and $C$ is the torsional stiffness, while $u_{10}$ and $u_{20}$ are the intrinsic curvatures and $u_{30}$ is the intrinsic torsion.

To these equations we apply the following zero-moment boundary conditions:
\begin{eqnarray}
x(0)=0, \quad\quad & \quad\quad x(1)=0, \nonumber \\
y(0)=0, \quad\quad & \quad\quad y(1)=0, \nonumber \\
z(0)=0, \quad\quad & \quad\quad F_z(1)=F_z, \nonumber \\
M_x(0)=0, \quad\quad & \quad\quad M_x(1)=0, \nonumber \\
M_y((0)=0, \quad\quad & \quad\quad M_y(1)=0, \nonumber \\
M_z(0)=0, \quad\quad & \quad\quad \bm{d}_1(1)\cdot\bm{i}=0,
\label{bcs}
\end{eqnarray}
where we have introduced components as follows: $\bm{r}=x\bm{i}+y\bm{j}+z\bm{k}$, $\bm{F}=F_x\bm{i}+F_y\bm{j}+F_z\bm{k}$, $\bm{M}=M_x\bm{i}+M_y\bm{j}+M_z\bm{k}$,
with $\{\bm{i},\bm{j},\bm{k}\}$ a fixed laboratory frame that has its origin at one end of the rod and $\bm{k}$ aligned with the (constant) force in the rod. Thus we impose the constraint that both ends of the rod remain on the line through $\bm{k}$.
Together with the six orthonormality conditions
\begin{equation}
\bm{d}_i(0)\cdot\bm{d}_j(0)=\delta_{ij}, \quad\quad j\geq i=1,2,3,
\end{equation}
where $\delta_{ij}$ is the usual Kronecker symbol, Eq.~(\ref{bcs}) gives a set of 18 boundary conditions for the 18 equations for the components of $\bm{r}$, $\bm{F}$, $\bm{M}$ and $\bm{d}_i$ ($i=1,2,3)$.

We solve this boundary-value problem using the continuation code AUTO \citep{Doedel98}, which allows for the tracking of solutions as parameters are varied and also detects bifurcations.

If we strongly penalise bending about the first material axis by pushing the ratio $A/B$ to infinity then the material frame will be locked onto the Frenet frame \citep{Kessler03}.
The corresponding equilibrium equations were derived by \citet{Starostin07}.

Before solving the equations, in the next section we consider the special case
where the unstressed rod is a multi-covered ring and the deformations are planar, described by Euler's elastica.

\section{Planar elastica with intrinsic curvature}

\subsection{Solving the equilibrium equations}

In this section we consider the planar shapes of a twist-free thin inextensible elastic rod. This degenerate planar case is important for the following analysis because it will help us better understand the behaviour of the low-pitch spring under tension. When left free of external forces and moments, the rod is assumed to take on the shape of a multicovered ring of radius $R_0$. Let it wind $n$ times
so that the intrinsic curvature becomes $\varkappa_0=1/R_0=2 \pi n$ (recall that the rod has unit length). Let $s_1$ and $s_2$ be the arclength coordinates of the ends of the rod so that $s_2=s_1+1$.

We are interested in the equilibrium configurations under an applied uniaxial non-vanishing tensile force $F$ at the ends of the rod. Equations (\ref{eq:bal})
in this case reduce to one equation for the tangent force $F_t=\bm{F}\cdot\bm{t}$:
\begin{equation}
({F_t}')^2 = (F_t+H)(F^2-F_t^2),
\label{eq2}
\end{equation}
where $H$ is a constant of integration and we have set $B=1$. Recall that the force vector is constant, hence $F^2 = \const$. The intrinsic curvature $\varkappa_0$ does not enter the equation, which is therefore that of the Euler elastica \citep{Love27}.

By introducing the angle $\vartheta$ between the fixed direction of the force and the tangent to the rod we can write $F_t = F\gamma=F\cos\vartheta$ and Eq.~(\ref{eq2}) takes the form
\begin{equation}
({\gamma}')^2 = (H+F\gamma)(1-\gamma^2)
\label{eq3}
\end{equation}
(cf.~Eq.~(2.8) in %(Starostin, 2004) 
\citep{Starostin04a}).
Integration of this equation gives
\begin{equation}
\gamma = 1 - 2 \sn^2(\Omega s),
\label{eq4}
\end{equation}
where $\Omega^2 = (H+F)/4$. For non-inflexional solutions the elliptic modulus $k$ is given by $k^2=2F/(H+F)$, where we require that $H>F$. For the force we can thus write $F=2 k^2\Omega^2$.

We introduce planar coordinates $(x,z)$ with $z$ chosen along the direction of the force. We may then write $x'=\pm\sqrt{1-\gamma^2}$, $z'=\gamma$, with $\gamma$ given by Eq.~(\ref{eq4}), and the centreline is found by integrating these equations (or they can be immediately obtained from Ilyukhin's equations reduced to the planar case %(Starostin, 2004)
\citep{Starostin04a}):
\begin{equation}
x=\frac{2}{k^2\Omega} \dn(\Omega s), \quad z=\left(1-\frac{2}{k^2}\right) s + \frac{2}{k^2\Omega} \E(\Omega s, k) ,
\label{eq5}
\end{equation}
where $\E(u, k) = \int_0^u \dn^2w \ \mbox{d} w$ is the incomplete elliptic integral of the second kind.

The curvature of the centreline is $\varkappa = k^2 \Omega^2 x$ (we note that always $x>0$).

The shape described by Eq.~(\ref{eq5}) is periodic with period $2\K(k)/\Omega$, where $\K(k)$ is the complete elliptic integral of the first kind.
Moreover, the centreline is symmetric with respect to the $x$-axis so that $x(s) = x(-s)$ and $z(s) =-z(-s)$.
The curvature monotonically decreases on the intervals $(2m\K(k)/\Omega,(2m+1)\K(k)/\Omega)$, $m \in \mathbb{Z}$, and monotonically increases on all other intervals.

We apply moment-free boundary conditions, i.e., $\varkappa(s_i) = \varkappa_0$ for $i=1,2$, which gives
\begin{equation}
\pi n = \Omega \dn(\Omega s_i), \quad\quad i=1,2.
\label{eq6}
\end{equation}
This implies that $x(s_1)=x(s_2)$ and that the force acts along the end-to-end line.

\subsection{Classification of solutions}

\subsubsection{Primary symmetric solutions}

We first consider symmetric shapes such that the midpoint of the centreline coincides with the extremum of the curvature, i.e.,
$(s_1+s_2)/2=m\K(k)/\Omega$, $m \in \mathbb{Z}$; it is a maximum for even $m$ and a minimum for odd $m$. Then the condition $s_2-s_1=1$ of fixed length transforms into
$s_i = m \K(k)/\Omega - (-1)^i/2$, $i=1,2$.
Substitution of the latter into Eq.~(\ref{eq6}) results in one of the two following conditions, depending on the parity of $m$:
\begin{eqnarray}
& \pi n = \Omega \dn{(\Omega/2)} & \quad \mbox{for $m$ even},  \nonumber \\
& \pi n = \Omega \sqrt{1-k^2} \nd{(\Omega/2)} & \quad \mbox{for $m$ odd}.
\label{eq7}
\end{eqnarray}
The end-to-end distance $\Delta z=z(s_1)-z(s_2)$ is computed from the second expression in Eq.~(\ref{eq5}) as
\begin{equation}
\Delta z = \left\{
\begin{array}{ll}
\frac{2}{k^2} - 1 - \frac{4}{k^2 \Omega} \E(\Omega/2,k) & \quad\mbox{for $m$ even}, \\
\frac{2}{k^2} - 1 - \frac{4}{k^2 \Omega} \E(\Omega/2,k)+\frac{4}{\Omega} \sn(\Omega/2)\cd(\Omega/2) & \quad\mbox{for $m$ odd}.
\end{array}
\right.
\label{eq8}
\end{equation}

\begin{figure}% FIGURE ================================================
\centering
\includegraphics[width=4in]{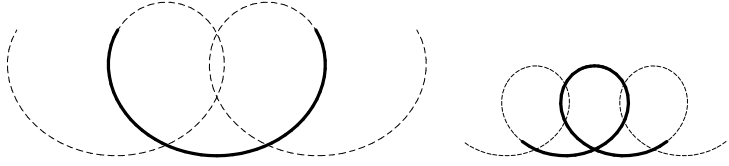}
\caption{\label{fig:u_a_shapes} Elastica configurations for $n=1$. {\it Left:} $\upsilon$-shape solution. {\it Right:} $\alpha$-shape solution. The end curvature in both cases is $\varkappa_0=2\pi n$. The elliptic moduli of the two solutions are the same. The force acts horizontally.}
\end{figure}

For given $k$ (and $m$) we consider Eq.~(\ref{eq7}) as an equation for $\Omega$.
For $k=0$, each of Eq.~(\ref{eq7}) has the unique solution $\Omega=\pi n$. By continuity in $k$ we can conclude that for given $n$ there exist only two branches of symmetric solutions through the undeformed initial state (one for even $m$ and one for odd $m$). Depending on the parity of $n-m$ these two types of solutions
have different shapes: if $n-m$ is even, then we call them {\it $\upsilon$-solutions}, otherwise we call them {\it $\alpha$-solutions} (see Fig.~\ref{fig:u_a_shapes}).
(This labelling is prompted by the superficial resemblance of the centreline configurations to the Greek letters.)
These two types may be thought of as opened ($\upsilon$) and tightened ($\alpha$) coils.

\begin{figure}% FIGURE ================================================
\centering
\includegraphics[height=2.9in]{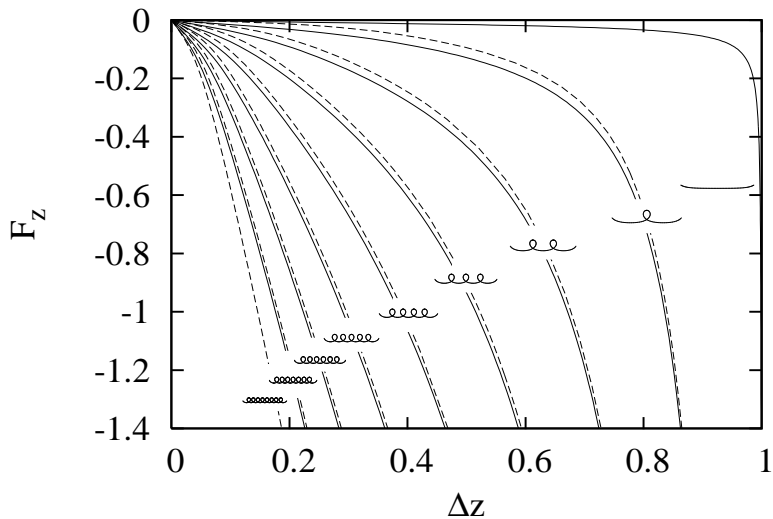}%  axes labels moved
\caption{\label{fig:f_z_sym} Force-extension curves for the rod with intrinsic curvature $\varkappa_0 = 2\pi n$. Symmetric branches passing through the origin are shown for $n=1,\ldots,8$ (increasing from right to left). Solid curves are for $\upsilon$-solutions, dashed curves for $\alpha$-solutions.
}
\end{figure}

Equation (\ref{eq7}) is solved numerically for increasing values of $k$.
Force-extension curves for both types of symmetric solutions are shown in Fig.~\ref{fig:f_z_sym}. Note that in this and all the subsequent figures we always show the {\it negative} external force that balances the {\it positive} internal force in the rod. The force is furthermore scaled by $(2\pi n)^2$ with $n=4$ for easy comparison with numerical results for the 4-turn helical spring in Section 4.

For the $\upsilon$-shapes we can approximate the solution for small $k \ll 1$
as $\Omega = \pi n + \frac{1}{128} \pi^3 n^3 k^6 + {\cal O}(k^8)$. Then $F=2 k^2 \Omega^2=2\pi^2n^2k^2+{\cal O}(k^8)$, while from  Eq.~(\ref{eq8}) for the extension $\Delta z$ we can write
\begin{equation}
\Delta z = \frac{3}{8} k^2 + \frac{15}{64} k^4 + \left(\frac{1225}{8192} - \frac{\pi^2 n^2}{96}\right) k^6 + {\cal O}(k^8) .
\label{eq9}
\end{equation}
For the $\alpha$-shapes the approximate solution of Eq.~(\ref{eq7}) is different:
$\Omega = \pi n \left(1 + \frac{1}{2} k^2 + \frac{3}{8} k^4 + \frac{40-\pi^2 n^2}{128} k^6 \right) + {\cal O}(k^8)$ and hence $F=2\pi^2 n^2 k^2 (1+k^2+k^4) + {\cal O}(k^8)$ and
\begin{equation}
\Delta z = \frac{3}{8} k^2 + \frac{9}{64} k^4 + \left(\frac{457}{8192} - \frac{\pi^2 n^2}{96}\right) k^6 + {\cal O}(k^8) .
\label{eq10}
\end{equation}
These approximations show that for small force and extension the slope of the force-extension curve of both types of solution is the same, i.e., the linear response is $F = \frac{16}{3} \pi^2 n^2 \Delta z$. Indeed the curve for one type of solution can be considered as a continuation of the curve of the other type of solution in the other direction (negative $\Delta z$ and positive $F_z$). For larger extensions the two curves diverge. For relatively large forces the curve
for the $\alpha$-solution corresponding to $n$ approaches the curve for the $\upsilon$-solution corresponding to $n+1$, as illustrated in Fig.~\ref{fig:f_z_sym}. Other symmetric solutions emerge at finite extension and force (resp., $k$), as we shall see below.

\subsubsection{Non-symmetric and secondary symmetric solutions}

\begin{figure}[t]% FIGURE ================================================
\centering
\includegraphics[width=2in]{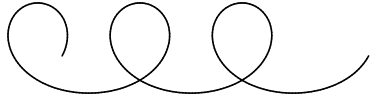}
\caption{\label{fig:theta_shapes} Example of a non-symmetric $\vartheta$-shape solution for $n=3$ and $j=3$. Both end curvatures are $\varkappa_0=2\pi n$. The force acts horizontally.}
\end{figure}

There are also non-symmetric configurations (we call them {\it $\vartheta$-shapes}) such that the length of the rod comprises an integer number of periods $j$, i.e., $2\K(k)j/\Omega = s_2-s_1 =1$, but the midpoint is not at maximum or minimum curvature (see Fig.~\ref{fig:theta_shapes}). Hence, $\Omega = 2 \K(k) j$ and the coordinate for the initial point of the rod $s_1$ can be found from the condition $\varkappa(s_1) = \varkappa_0 = 2\pi n$, which gives
\begin{equation}
\pi n = 2 \K(k) j \dn (2 \K(k) j s_1) .
\label{eq81}
\end{equation}

\begin{figure}[b]% FIGURE ================================================
\centering
\includegraphics[width=4.5in]{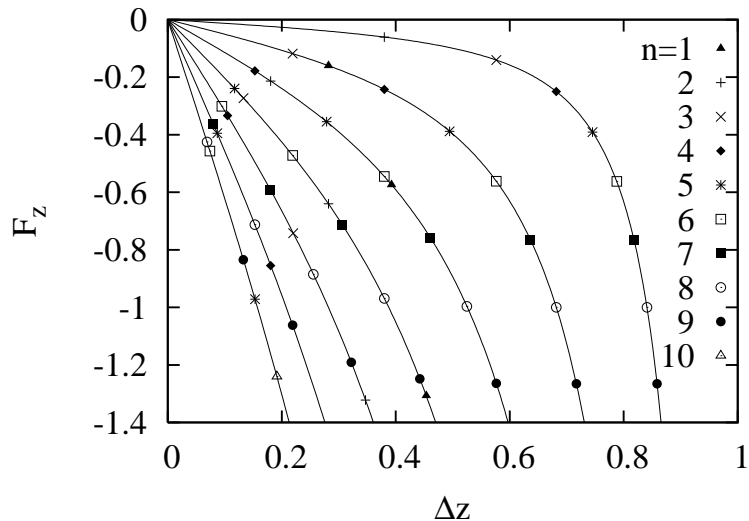}
\caption{\label{fig:f_z_nonsym} Force-extension graphs for non-symmetric configurations with number of periods $j=1,\ldots,7$ (increasing from right to left). Bifurcation points are shown for various values of $n$.}
\end{figure}

The extension of non-symmetric solutions is computed from Eq.~(\ref{eq5}) with $\Omega = 2 \K(k) j$:
\begin{equation}
\Delta z = \frac{2}{k^2}\left(1- \frac{\E(k)}{\K(k)}\right) -1 ,
\label{eq82}
\end{equation}
where $\E(k)$ is the complete elliptic integral of the second kind.
The force is given by $F=8k^2\K(k)^2 j^2$. Force vs. extension curves are shown in 
Fig.~\ref{fig:f_z_nonsym} for $j=1,\ldots,7$. 

Note that the loading curves in Fig.~\ref{fig:f_z_nonsym} do not themselves depend on the intrinsic curvature, i.e., on the number of initial coils $n$, but their parametrisation by $k$ does, as is seen in Eq.~(\ref{eq81}). For $k=0$ this equation has a solution only for $j=n$ (then $\Omega = \pi n$) and, by continuity, this is also true for small $k$. For such $k$ we can write, from Eq.~(\ref{eq82}),
\begin{equation}
\Delta z = k^2 \left(\frac{1}{8} + \frac{1}{16} k^2 + \frac{41}{1024} k^4 + {\cal O}(k^6) \right).
\label{eq11}
\end{equation}
Developing the force into the series in $k$, we have, with $j=n$,
\begin{equation}
F = \pi^2 n^2 k^2 \left(2 + k^2 + \frac{11}{16} k^4 + {\cal O}(k^6) \right).
\label{eq12}
\end{equation}
Combining both expressions we find the linear approximation
$F=16\pi^2 n^2 \Delta z$, valid for small deformations. We see that the non-symmetric solution requires a force three times that of the symmetric one.

As $k$ increases, other roots of Eq.~(\ref{eq81}) appear. A new root comes into existence each time the Jacobi elliptic function $\dn$ reaches a maximum or minimum, i.e., for $s_1 = \frac{m}{2j}$ (even $m$ corresponds to a maximum, odd $m$ to a minimum). A maximum can only occur for $j < n$, because $\frac{n}{j}= \frac{2\mathrm{K}(k)}{\pi} >1$ for $k>0$, while a minimum can only occur for $j > n$, because $\frac{n}{j}= \frac{2\mathrm{K}(k)\sqrt{1-k^2}}{\pi} <1$ for $k>0$.

Thus, for given actual number of periods $j$ the corresponding curve in Fig.~\ref{fig:f_z_nonsym} represents the shapes for various intrinsic curvatures. For $n=j$ the entire curve is realisable; otherwise, only part of it is realisable
with some offset from the origin. The offset increases with the difference between the number of initial coils $n$ and actual periods $j$.

\begin{figure}% FIGURE ================================================
\centering
\includegraphics[width=5.5in]{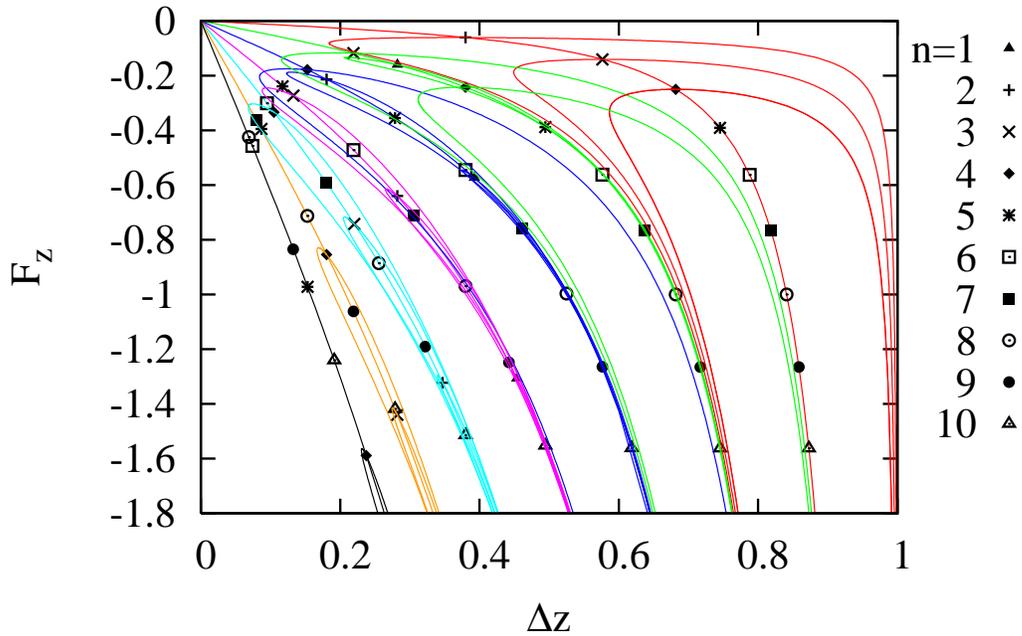} 
\caption{\label{fig:f_z_nonsym_n} Force-extension curves for a rod with intrinsic curvature $\varkappa_0 = 2\pi n$ with $n=1,\ldots,10$. Non-symmetric and some symmetric branches that do not pass through the origin are shown for $j=1,\ldots,7$. Bifurcation points are shown for various values of $n$.}
\end{figure}

\begin{figure}[b]% FIGURE ================================================
\centering
\includegraphics[width=2in]{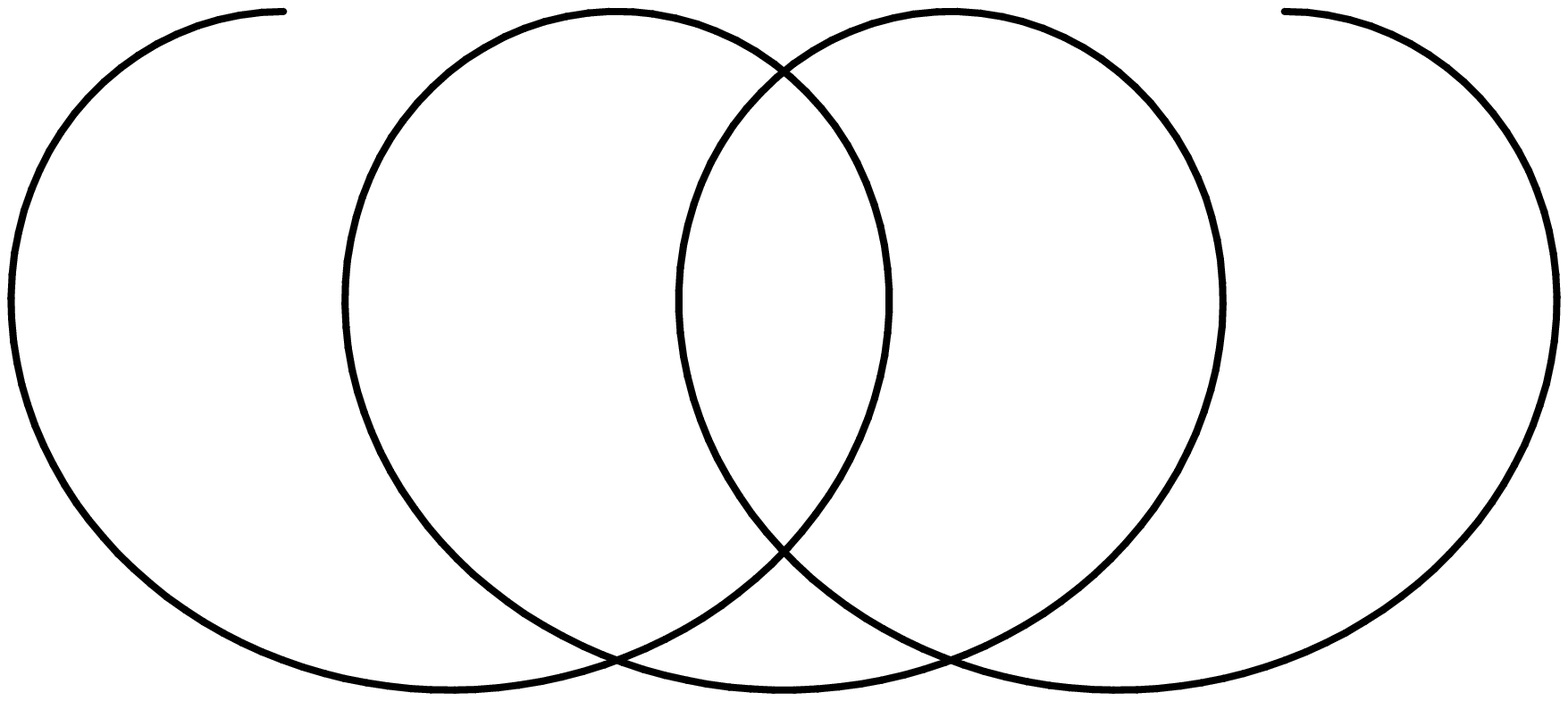}  \quad \includegraphics[width=2in]{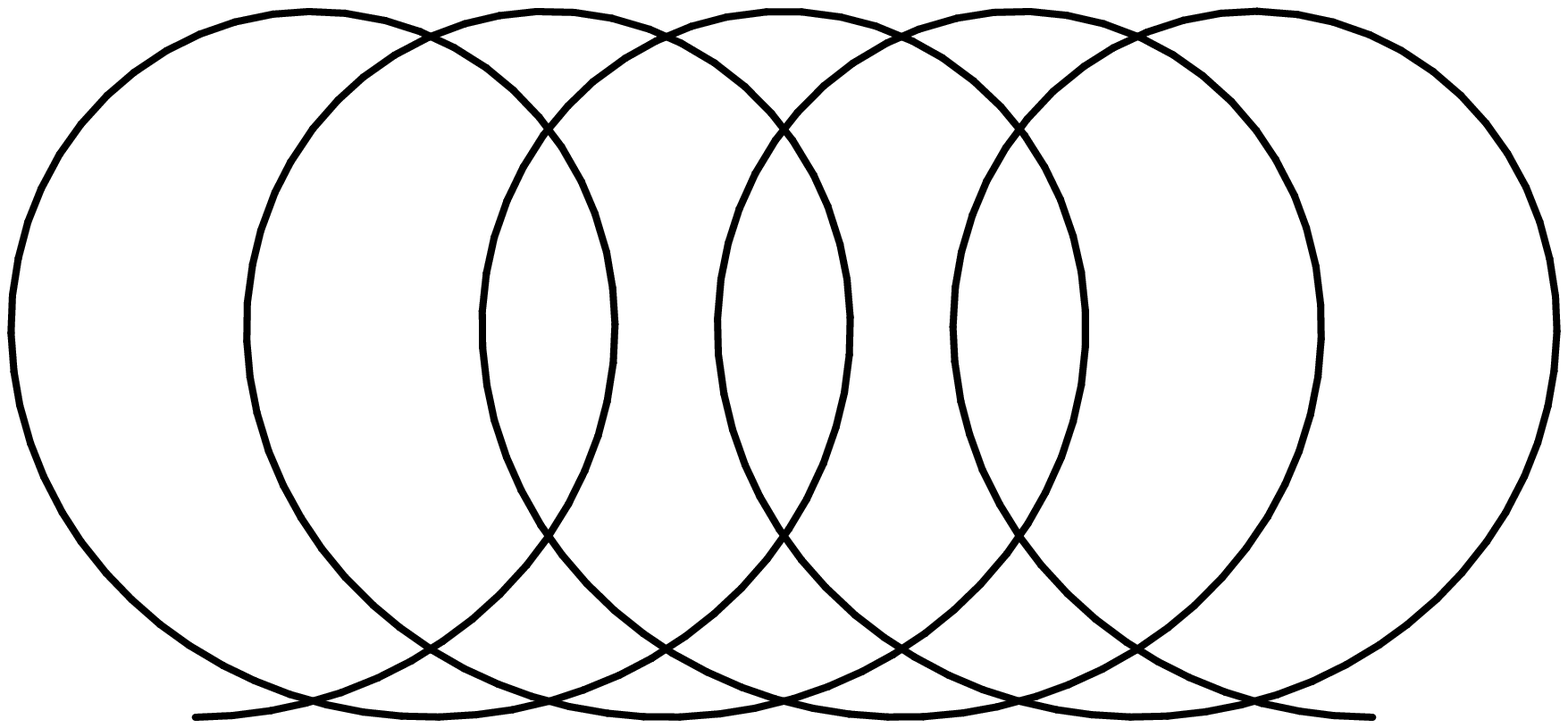}
\caption{\label{fig:bif_shapes} Elastica configurations at intersection points of symmetric and non-symmetric branches for $n=4$. {\it Left:} $\upsilon$-shape solution for $j=3$. {\it Right:} $\alpha$-shape solution for $j=5$. The force acts horizontally.}
\end{figure}

The values of $k$ where new roots emerge correspond to configurations at the intersection of the non-symmetric branch with {\it secondary symmetric} branches that do not pass through the origin (see Fig.~\ref{fig:f_z_nonsym_n}). For $j < n$, they are $\upsilon$-shapes with their ends at points of maximum curvature, while for $j > n$ they are $\alpha$-shapes with their ends at points of minimum curvature (see Fig.~\ref{fig:bif_shapes}). In both cases the end tangent vectors are along the end-to-end distance and the $z$-axis.

\begin{figure}% FIGURE ================================================
\centering
\includegraphics[height=3in]{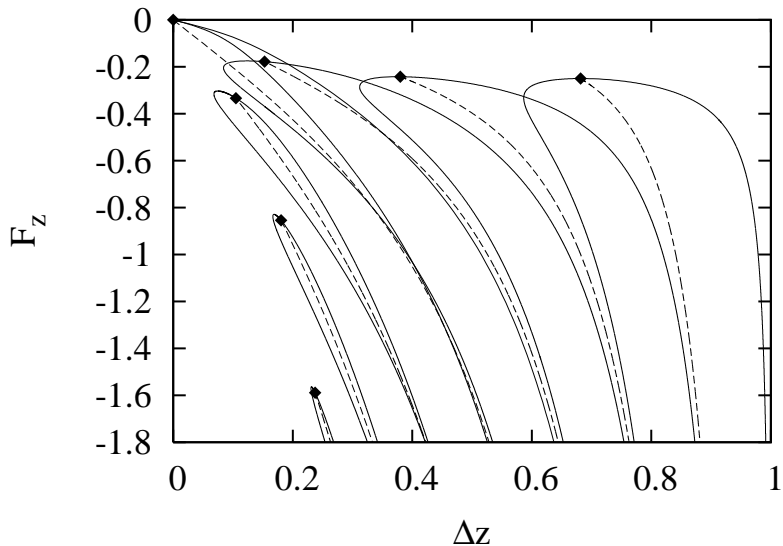}
\caption{Primary and secondary symmetric (solid) as well as non-symmetric (dashed) branches for $n=4$ and $j=1,\ldots,7$ (increasing from right to left).}
\label{fig:f_z_nonsym_symm_n4}
\end{figure}

If we fix the intrinsic curvature, i.e., the number $n$, then the set of the non-symmetric solutions consists of 1) one complete curve $j=n$ that begins in the origin and 2) parts of all other curves starting from the bifurcation points. There is only one bifurcation point on each $j$-curve and the symmetric branch is born here. The solutions on these branches are computed from Eqs.~(\ref{eq7}) and (\ref{eq8}) with $m=j$ for $j<n$ and $m=j+1$ for $j>n$. Fig.~\ref{fig:f_z_nonsym_symm_n4} shows non-symmetric and bifurcating symmetric branches for $n=4$. It is convenient to label the symmetric branches with the number $j$. As the applied force increases, the symmetric $j$-branches for $j>n$ ($\alpha$-shapes) converge to non-symmetric branches with the same $j$, while the 
branches for $j<n$ ($\upsilon$-shapes) approach their two neighbouring non-symmetric curves with $j\pm1$, i.e., they either lose or gain one coil.
Meanwhile, the non-symmetric $j$-branch approaches the primary symmetric $n$-branch with $n=j+1$.

Physically, solutions on the bifurcating symmetric branches, which do not go through the origin, can only be obtained by uncoiling the $n$-loop and recoiling into a shape with $j$ loops. They cannot be held at zero force.

Summing up, we see that for given intrinsic curvature there exist 1) only three branches (symmetric $\alpha$- and $\upsilon$-shapes and a non-symmetric $\vartheta$-shape) that contain the relaxed state, 2) $n-1$ associated pairs of opened $\vartheta$- and $\upsilon$-solutions, 3) a countably infinite number of tightened $\vartheta$- and $\alpha$-solutions. Clearly, no branch of opened or tightened shapes can approach the relaxed state of the $n$-covered ring.

%%%%%%%%%%%%%%%%%%%%%%%%%%%%%%%%%%%%%%%%%%%%%%%%%%%%%%%%%%%%%%%%%%%%%%%%%%%%%%%%%%%%%%%%%%%%%%%%%%%%%%%%%%%%%%%%%%%%%%%%%%%%

\section{Pulling a helical spring -- Numerical results}

Having classified all planar solutions we now turn to the three-dimensional deformations of our helical spring.

\begin{figure}% FIGURE ================================================
\centering
\includegraphics[width=4.5in]{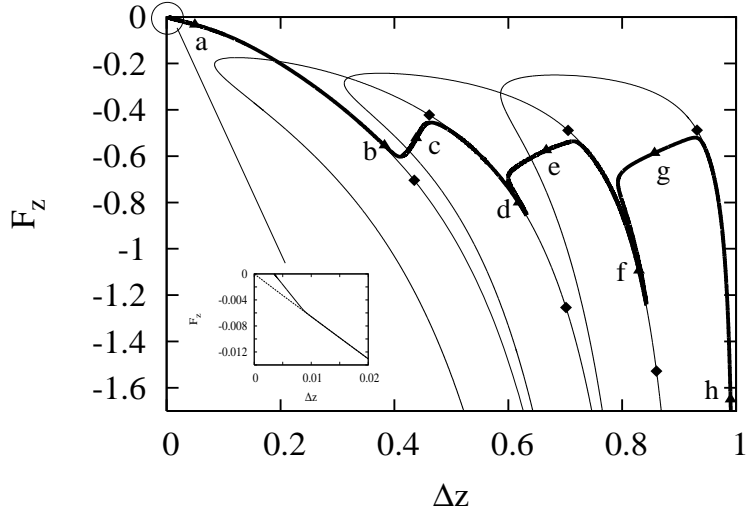} 
\caption{Force-extension curve for a low-pitch helix with $n=4$ (thick curve) and approximating planar elastica curves. The diamonds indicate points of out-of-plane instability of the elastica. The enlargement reveals a sharp transition from 3D to 2D behaviour. Labels `a' to `h' refer to solutions displayed in Fig.~\ref{shapes}. ($\theta_0=89.8^\circ$, $A/B=10$, $\nu=0.05$.)}
\label{spring}
\end{figure}

Figure \ref{spring} gives the force-extension curve for an unstressed helix of unit length with 4 turns (i.e., $n=4$) and helical angle $\theta_0=89.8^\circ$ (i.e., a pitch angle of 0.2$^\circ$), subject to the above boundary conditions. This helix has radius $R_0=\sin\theta_0/(2\pi n)$, curvature $\varkappa_0=2\pi n\sin\theta_0$
and geometrical torsion $\tau_0=2\pi n\cos\theta_0$. The elastic constants are fixed such that $A/B=10$ and $C/B=1.7316$, and we set $u_{10}=0$, $u_{20}=\varkappa_0$ and $u_{30}=\tau_0$. These values imply that the smaller bending stiffness is along the axis of intrinsic curvature. The value of $C/B$ corresponds to a Poisson ratio of $\nu=0.05$ if we assume the elastic constants to be at the boundary of the admissible region for homogeneous cross-sections of arbitrary geometry, as given by the Nikolai inequality
\citep{Ilyukhin79}%
\begin{equation}
C\leq\frac{2AB}{(1+\nu)(A+B)}\cdot
\end{equation}
(Here equality holds only for elliptical cross-sections.)
All forces are scaled by $(2\pi n)^2B$.

\begin{figure}% FIGURE ================================================
\centering
\includegraphics[width=5.5in]{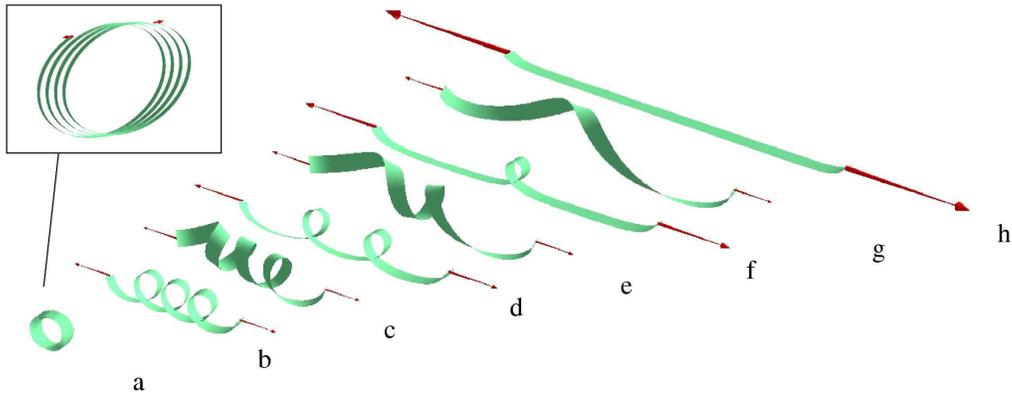}
\caption{Solution shapes along the curve in Fig.~\ref{spring}. Labels loosely correspond to those in Fig.~\ref{photos}. Note the shearing behaviour at small force (inset).
($n=4$, $\theta_0=89.8^\circ$.)}
\label{shapes}
\end{figure}

The figure shows that after a short stretch, during which the helix turns and shears, the force-extension curve very closely follows the (primary symmetric) planar elastica curve until the elastica's out-of-plane instability is approached. The curve then sharply veers away from the $j=4$ elastica curve and solutions become three-dimensional. This is illustrated by the shapes depicted in Fig.~\ref{shapes}, which bear good resemblance to the (identically labelled) actual strip shapes in Fig.~\ref{photos}. After this excursion the curve lands on the 
(secondary symmetric) $j=3$ branch, its solutions having shed one loop. Two more such unlooping manoeuvres occur while the force-extension curve jumps from the $j=3$ to the $j=2$ and finally to the $j=1$ branch. Associated with these jumps is hysteresis behaviour, i.e, the jumps occur at different points if the extension $\Delta z$ is decreased rather than increased. It is worth noting that all elastica curves visited by the spring curve are $\upsilon$-branches.
The small pitch angle acts as an imperfection causing a rounding off of the elastica bifurcation diagram which would have a branch of three-dimensional solutions connecting the pitchfork bifurcations indicated by diamonds. Note that the location of these diamonds, but not the curves, depends on the torsional stiffness $C$.
Clearly, for the nearly planar solutions not to have self-contact, which is not modelled, we have to assume the strip to be sufficiently narrow.

\begin{figure}% FIGURE ================================================
\centering
\includegraphics[width=4.5in]{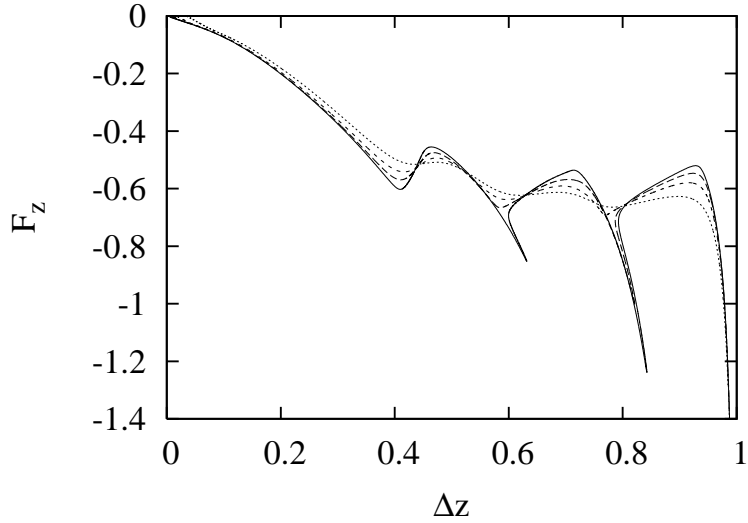} 
\caption{Force-extension curves for different helical angles: $\theta_0=89.8^\circ$ (solid), 89.5$^\circ$, 89$^\circ$ and 88$^\circ$ ($n=4$, $A/B=10$, $\nu=0.05$).}
\label{th0}
\end{figure}

\begin{figure}% FIGURE ================================================
\centering
\includegraphics[width=4.5in]{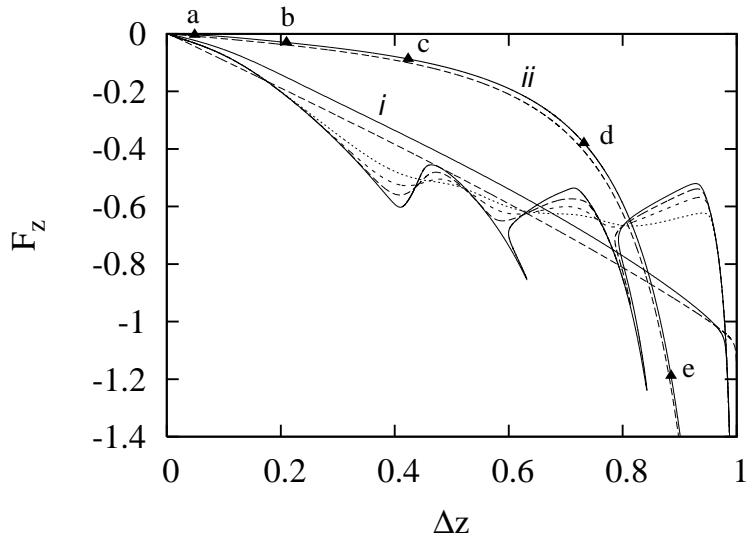} 
\caption{Force-extension curves for different anisotropies: $A/B=10$ (solid), 8, 6 and 4. Also shown are curves for isotropic ($A/B=1$, $i$) and slinky ($A/B=1/10$, $ii$) rods (solid), together with approximations (dashed) obtained by assuming an exact helical shape throughout. ($n=4$, $\theta_0=89.8^\circ$, $\nu=0.05$.)}
\label{aniso}
\end{figure}

\begin{figure}% FIGURE ================================================
\centering
\includegraphics[width=4.5in]{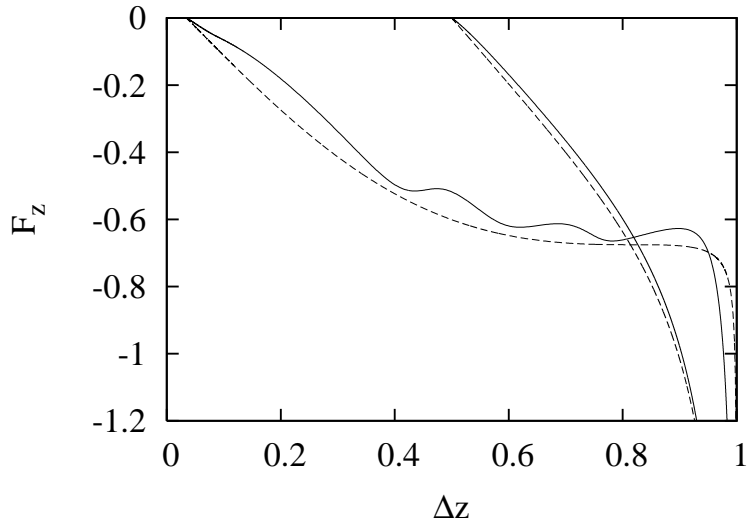} 
\caption{Force-extension curves (solid) for a helical spring at $\theta_0=88^\circ$ ($\Delta z_0=\cos\theta_0=0.03490$) and $60^\circ$ ($\Delta z_0=\cos\theta_0=0.5$) compared against curves (dashed) obtained by assuming an exact helical shape throughout ($n=4$, $A/B=10$, $\nu=0.05$).}
\label{helix}
\end{figure}

Figures \ref{th0} and \ref{aniso} show the dependence of this unlooping behaviour on $\theta_0$ and anisotropy $A/B$. It is seen that the unlooping cascade requires, in addition to both a small pitch and a small $\nu$, a relatively large anisotropy, otherwise the force-extension behaviour resembles that of a helix, which gradually unwinds under an applied tension. To illustrate this, in Fig.~\ref{helix} we compare force-extension curves for $\theta_0=88^\circ$ and $60^\circ$ with curves obtained by assuming an exact helical shape for the rod throughout the tension test. For this case, which would require special boundary conditions, exact expressions can be derived. These are given in the Appendix. We note that for a pitch angle as small as $2^\circ$ ($\theta_0=88^\circ$) the agreement with the helical curve is already reasonably good, although the latter fails to account for multiple local minima. At relatively large pitch angles (e.g., $\theta_0=60^\circ$) the hysteresis behaviour disappears.

\begin{figure}% FIGURE ================================================
\centering
\includegraphics[width=5.5in]{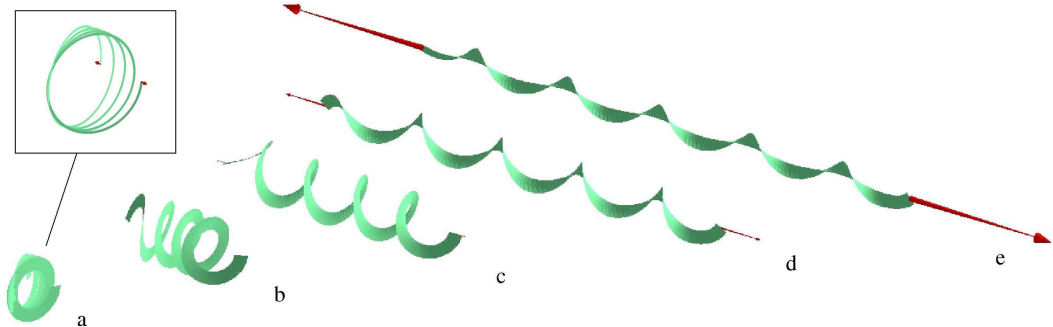}
\caption{Slinky solution shapes along the curve labelled $ii$ in Fig.~\ref{aniso}. 
The inset shows shape `a' for a narrower strip to reveal the characteristic slinky response. ($n=4$, $\theta_0=89.8^\circ$.)}
\label{slinky_shapes}
\end{figure}

For comparison, in Fig.~\ref{aniso} we also present force-extension curves for
the slinky spring, whose axis of least bending stiffness is at right angles to the axis of intrinsic curvature ($A/B=1/10$), as well as for the isotropic rod ($A/B=1$). In both cases no hysteresis is found and the curves are well approximated by helical curves (included in dashed lines). Slinky shapes are shown in Fig.~\ref{slinky_shapes} revealing that the slinky spring is pulled out into an almost perfectly helical shape. Interestingly, rather than unwinding, the spring overwinds under increasing force.

\section{Concluding remarks}

We have analysed the problem of successive unlooping of a helical spring. Springs of sufficiently low pitch made of material of sufficiently large ratio of torsional to bending stiffness are found not simply to unwind when pulled, but rather to form hockles that successively pop out. The force-extension curves are non-monotonic with large parts well approximated by curves for the planar Euler elastica and connected by brief three-dimensional excursions. Similar complicated force-extension behaviour was found in the helical shell model proposed by \citet{Starostin08a}. In that study we showed that the spring force-extension curves can be approximated by (primary) elastica curves (analogous to those in Fig.~\ref{fig:f_z_sym}). In the present paper we have shown that the force-extension curves are even better approximated by the {\it secondary} elastica curves that bifurcate from non-symmetric branches (as in Fig.~\ref{fig:f_z_nonsym_symm_n4}).

A low-pitch helix can be interpreted as a `perturbation' of a planar circular spring. Our analysis gives an unfolding of the bifurcation diagram of the perfect planar elastica. The unlooping scenario in the unfolding describes the tensile response of an imperfect planar spring.

Our study has been restricted to (quasi-)statical solutions. In practice the three-dimensional excursions will be rapid and would require dynamical modelling, as for instance in \citep{Goyal05}. Our work also paid no attention to self-contacts. This is justified by the fact that no shapes we computed cross themselves (the only exception is the slinky shape at very small force). Thus choosing cross-sections small enough excludes self-contact.

By choosing our torsional stiffness to satisfy the Nikolai condition as an equality we take the maximum possible value for any solid cross-section, given the bending stiffnesses $A$ and $B$ and the Poisson ratio $\nu$. Even so, we need quite a small value of $\nu$ to observe the unlooping cascade described in the previous sections. However, moving away from solid cross-sections, materials (e.g., biopolymers) are now well-known that have a relatively large torsional stiffness corresponding to a small or even negative effective Poisson ratio. For instance, double-stranded DNA is exceptionally stiff in torsion \citep{Oroszi06}. It is also worth pointing out that the inextensible helical shell model considered by \citet{Starostin08a} would exhibit the unlooping cascade for a larger range of $\nu$ values because of its higher intrinsic torsional stiffness.

Our results may be useful for the design of mechanisms for the deployment of ribbon-like structures. The nonlinear properties of the elastic response of helical strips could be exploited in mechanical sensors and actuators at the micro- or nanoscale, where low-pitch springs are being considered.

\appendix

\section{Force-extension curves for helical solutions}

Helical curves are characterised by constant curvature and torsion. Since the stiffness ratio $A/B$ is generally taken relatively large in this paper ($A/B=10$ in Figs.~\ref{spring} and \ref{helix}) bending of the rod will predominantly be about $\bm{d}_2$, the direction of intrinsic curvature. Thus we seek helical solutions of the equations in Section 2 with $u_1=0$, $u_2=\varkappa=\mbox{const.}$, $u_3=\tau=\mbox{const.}$ It is convenient to introduce the helical angle $\theta$ (cf.~Fig.~\ref{fig:spring}) such that $\tan\theta=\varkappa/\tau$. Force and extension for a unit-length helix are then parametrised in terms of $\theta$ as follows:
\begin{eqnarray}
&& F = -\frac{BC\sin^2\theta_0(B \sin\theta_0\sin\theta + C \cos\theta_0\sin\theta) \sin(\theta - \theta_0)}{R_0^2 \sin\theta (B\sin^2\theta + C \cos^2\theta)^2}, \\
&& \Delta z=\cos\theta,
\end{eqnarray}
where $R_0=\sin^2\theta_0/\varkappa_0$ is the radius of the undeformed helix and $z$ is a coordinate along its axis. Note that the force is now pointing along the axis of the helix and not necessarily along the end-to-end vector. The radius of the deformed helix is given by
\begin{equation}
R = R_0 \frac{\sin\theta(B\sin^2\theta + C \cos^2\theta)}{\sin\theta_0 (B\sin\theta \sin\theta_0 + C \cos\theta \cos\theta_0)}\cdot
\end{equation}

In Fig.~\ref{helix} force-extension curves for these helical solutions are included
for comparison with the force-extension curves for a pulled helix.

%%\begin{thebibliography}{}

% \bibitem[Names(Year)]{label} or \bibitem[Names(Year)Long names]{label}.
% (\harvarditem{Name}{Year}{label} is also supported.)

%%\bibitem[]{}

%%\end{thebibliography}
\bibliography{}

\end{document}